\documentclass[10pt]{iopart}
\usepackage{iopams}
\usepackage{epsfig}

\newcommand{\bra}[1]{{\left\langle #1 \right|}}
\newcommand{\ket}[1]{{\left| #1 \right\rangle}}

\begin{document}
\title{Entanglement for a two-parameter class of states in $2\otimes n$ quantum system}

\author{
Dong Pyo Chi\dag and Soojoon Lee\ddag}
\address{\dag\ School of Mathematical Sciences,
Seoul National University, Seoul 151-742, Korea}
\address{\ddag\ School of Computational Sciences,
Korea Institute for Advanced Study, Seoul 130-722, Korea}
\eads{\mailto{dpchi@math.snu.ac.kr}, \mailto{level@kias.re.kr}}

\begin{abstract}
We exhibit a two-parameter class of states $\rho_{(\alpha,\gamma)}$,
in $2\otimes n$ quantum system for $n\ge 3$,
which can be obtained from an arbitrary state
by means of local quantum operations and classical communication,
and which are invariant under all bilateral 
operations 
on $2\otimes n$ quantum system.
We calculate the negativity of $\rho_{(\alpha,\gamma)}$,
and a lower bound and a tight upper bound on its entanglement of formation.
It follows from this calculation
that the entanglement of formation of $\rho_{(\alpha,\gamma)}$
cannot exceed its negativity.
\end{abstract}

\section{Introduction}


Entanglement is one of the most important resources
for quantum communication and information processing
including quantum cryptography, teleportation, and superdense coding.
On this account,
the research on entanglement has considerably been developed
and has improved the quantum information science in recent years.
In particular, the quest for proper measures of entanglement
has received a great deal of attention,
and several measures of entanglement, such as the negativity and the entanglement of formation,
have been proposed \cite{BDSW,Wootters,Horodecki,Wootters2,VDC}.

Peres-Horodeckis' criterion for separability \cite{Peres,Horodeckis1}
leads a natural entanglement measure, called the {\em negativity} $\mathcal{N}$,
defined by
\begin{equation}
\mathcal{N}(\rho)={\|\rho^{T_B}\|_1-1},\label{eq:negativity}
\end{equation}
where $\rho^{T_B}$ is the partial transpose of a state $\rho$
in a Hilbert space $\mathcal{H}_A\otimes\mathcal{H}_B$
and $\|\cdot\|_1$ is the trace norm.
Vidal and Werner~\cite{VidalW} defined the negativity of a state $\rho$ as
$(\|\rho^{T_B}\|_1-1)/2$,
which corresponds to the absolute value of the sum of negative eigenvalues of $\rho^{T_B}$,
and which vanishes for separable states.
For a maximally entangled pure state such as one of the Bell states,
this quantity is strictly less than one.
In order for any maximally entangled pure state in $2\otimes n$ quantum system
to have the negativity one,
it must be defined as \Eref{eq:negativity}.

We note that the negativity is an entanglement monotone~\cite{Vidal}
under local quantum operations and classical communication (LOCC) \cite{VidalW}, and
that it is a measure of entanglement which can be computed effectively for any state.
However, although the positivity of the partial transpose is a necessary and sufficient condition
for nondistillability in $2\otimes n$ quantum system \cite{Horodecki1,DCLB},
there exist entangled states with positive partial transposition (PPT) in any bipartite system
except for in $2\otimes 2$ and $2\otimes 3$ quantum systems \cite{Horodecki1,Horodeckis2},
and hence it is not sufficient for the negativity to be a good measure of entanglement
even in $2\otimes n$ quantum system.

In this paper we consider $2\otimes n$ quantum systems for $n\ge 3$,
and exhibit a two-parameter class of states
in $2\otimes n$ quantum system,
which can be obtained from an arbitrary state
by means of LOCC
and are invariant under all unitary operations of the form $U\otimes U$
on $2\otimes n$ quantum system.\footnote[7]{
Let $U(k)$ be the group of all unitary operators on a $k$-dimensional Hilbert space,
and $\left\{\ket{0}_A, \ket{1}_A\right\}$ and $\left\{\ket{0}_B,\ket{1}_B, \ldots, \ket{n-1}_B\right\}$
be bases of $\mathcal{H}_A$ and $\mathcal{H}_B$ respectively.
For convenience,
we now identify a unitary operator $U_A\in\mathrm{U}(2)$
with $U_B\in\mathrm{U}(n)$
if for $j=0,1$,
$U_A\ket{j}_A=a_j\ket{0}_A+b_j\ket{1}_A$ and
$U_B\ket{j}_B=a_j\ket{0}_B+b_j\ket{1}_B$.
 For $0<m<n$,
we let
\begin{equation}
\mathrm{G}(m,n)
=\left\{U\in\mathrm{U}(n): U\left(\mathcal{H}_{m}\right)=\mathcal{H}_{m},
U\left(\mathcal{H}_{m}^\perp\right)=\mathcal{H}_{m}^\perp \right\},
\label{eq:G(m,n)}
\end{equation}
where $\mathcal{H}_{m}$ is a subspace of $\mathcal{H}_B$
generated by $\ket{0}_B,\ket{1}_B,\ldots, \ket{m-1}_B$,
and $\mathcal{H}_{m}^\perp$ is the orthogonal complement of $\mathcal{H}_{m}$ in $\mathcal{H}_B$.
Then $\mathrm{G}(2,n)$ is a subgroup of $\mathrm{U}(n)$, and
if $U$ is a unitary operator in $\mathrm{G}(2,n)$ then
it is compatible to write a unitary operator of the form $U\otimes U$ on $2\otimes n$ quantum system.
}
We show that a state in the two-parameter class has a PPT if and only if
the state is separable,
so that the negativity can be a measure to quantify the amount of entanglement of states
in the class.
We remark that a finite dimensional truncation of
a single two-level atom interacting with a single-mode quantized field \cite{SZ},
can be regarded as a $2\otimes n$ quantum system.

The {\em entanglement of formation} is defined to be the convex-roof extension of the pure-state entanglement,
that is,
the minimum average of the pure-state entanglement over all ensemble decompositions of a given state $\rho$,
\begin{equation}
E_f(\rho)=\min_{\sum_{k}p_k \ket{\psi_k}\bra{\psi_k}=\rho}
\sum_{k}p_k E(\ket{\psi_k}).
\label{eq:Eof}
\end{equation}
Here, the pure-state entanglement $E$ is defined as the entropy of subsystem $A$,
$E(\ket{\psi})=S({\rm tr}_B(\ket{\psi}\bra{\psi}))$, $S$ is the von Neumann entropy.

Since the concept of the pure-state entanglement $E$
has most widely been accepted
as an entanglement measure for pure states
and the entanglement of formation is its natural extension as seen in \Eref{eq:Eof},
the entanglement of formation is one of the best measures for bipartite quantum systems
which have been known so far.
Nevertheless, there is no known explicit formula
for the entanglement of formation of states in a general quantum system
except for states in $2\otimes 2$ quantum system \cite{Wootters,HW},
the isotropic states and the Werner states in $n\otimes n$ quantum system \cite{Wootters2,TV,VollbrechtW},
and states of the specific form \cite{Wootters2,VDC}.
For $2\otimes n$ quantum system, only a lower bound on the entanglement of formation is given by
decomposing a $2\otimes n$ dimensional Hilbert space into many $2 \otimes 2$ dimensional subspaces
\cite{CLLH,Gerjuoy}.

In this paper we present a lower bound and a tight upper bound
on the entanglement of formation for the two-parameter class of states.
The explicit calculations of the negativity and bounds on the entanglement of formation
show that the entanglement of formation of any state in the two-parameter class
cannot exceed its negativity.

This paper is organized as follows:
In \Sref{Sec:2parameter} we exhibit a two-parameter class of states and
a procedure involving only LOCC
which transform an arbitrary state 
into one of states in the class,
and show that the states in the class are invariant
under all unitary operations of the form $U\otimes U$. 
In \Sref{Sec:Entanglement} we explicitly calculate
the negativity, and a lower bound and a tight upper bound on the entanglement of formation
for the two-parameter class,
and compare its negativity with the entanglement of formation.
Finally, in \Sref{Sec:Conclusion} we summarize our results
and discuss a generalization of the two-parameter class into a higher dimensional system.

\section{A two-parameter class of states in $2\otimes n$ quantum system}\label{Sec:2parameter}

We consider the following class of states with two real parameters $\alpha$ and $\gamma$
in $2\otimes n$ quantum system:
\begin{eqnarray}
\rho_{(\alpha,\gamma)}&=&\alpha \sum_{i=0}^1\sum_{j=2}^{n-1}\ket{ij}\bra{ij} \nonumber \\
&&+\beta \left(\ket{\phi^+}\bra{\phi^+}+\ket{\phi^-}\bra{\phi^-}+\ket{\psi^+}\bra{\psi^+}\right) \nonumber \\
&&+\gamma \ket{\psi^-}\bra{\psi^-}\label{eq:2parameter}
\end{eqnarray}
where $\left\{\ket{ij}: i=0,1, j=0,1,\ldots, n-1\right\}$ is an orthonormal basis for $2\otimes n$ quantum system,
\begin{eqnarray}
\ket{\phi^{\pm}}&=&\frac{1}{\sqrt{2}}\left(\ket{00}\pm\ket{11}\right),
\\
\ket{\psi^{\pm}}&=&\frac{1}{\sqrt{2}}\left(\ket{01}\pm\ket{10}\right),
\end{eqnarray}
and
the parameter $\beta$ is dependent on $\alpha$ and $\gamma$ by the unit trace condition,
\begin{equation}
2(n-2)\alpha+3\beta+\gamma=1.\label{eq:unit_trace_cond}
\end{equation}
 From the unit trace condition in \Eref{eq:unit_trace_cond}
we can readily obtain the domain for the parameters $\alpha$ and $\gamma$,
$0\le \alpha \le 1/2(n-2)$ and $0\le \gamma\le 1$.
We note that the states of the form $\rho_{(0,\gamma)}$ for $0\le \gamma \le 1$ are equal to
Werner states~\cite{Werner}
in $2\otimes 2$ quantum system,
that the states are entangled and distillable if and only if $1/2<\gamma\le 1$,
and that for $1/2\le \gamma\le 1$,
\begin{eqnarray}
\mathcal{N}(\rho_{(0,\gamma)})&=&2\gamma-1,\nonumber\\
E_f(\rho_{(0,\gamma)})&=& h\left(\frac{1+\sqrt{1-\mathcal{N}(\rho_{(0,\gamma)})^2}}{2}\right)
= h\left(\frac{1}{2}+\sqrt{\gamma\cdot(1-\gamma)}\right) \nonumber
\end{eqnarray}
where $h$ is the binary Shannon entropy.

We are going to show now that an arbitrary state $\rho$ in $2\otimes n$ quantum system
can be transformed to a state of the form $\rho_{(\alpha,\gamma)}$ in \Eref{eq:2parameter}
by LOCC.
In other words,
we will show that there exist unitary operators $U_k$ and probabilities $p_k$
such that
\begin{equation}
\sum_k p_k (U_k\otimes U_k) \rho
(U_k^{\dagger}\otimes U_k^{\dagger})=\rho_{(\alpha,\gamma)}
\label{eq:LOCC}
\end{equation}
for some $0\le \alpha \le 1/2(n-2)$ and $0\le \gamma\le 1$,
using the method similar to those
presented by Bennett {\it et al.} \cite{BDSW} and D\"{u}r {\it et al.} \cite{DCLB}.


We define the operation $U_{\theta}$ as
\[
U_{\theta}:\ket{j}\mapsto (e^{\mathrm{i}\theta})^j \ket{j},
\]
where $\mathrm{i}=\sqrt{-1}$.
We first perform $U_{\pi}\otimes U_{\pi}$ with probability $1/2$,
while with probability $1/2$ no operation is performed,
that is,
\begin{equation}
\frac{1}{2}\left(U_{\pi}\otimes U_{\pi}\right) \rho (U_{\pi}^{\dagger}\otimes U_{\pi}^{\dagger})
+\frac{1}{2}\rho.\label{eq:U_piU_pi}
\end{equation}
Let us now define the operation $U_k$ by $U_k:\ket{j}\mapsto (-1)^{\delta_{j,k}}\ket{j}$ for $k=2,3,\ldots,n-1$,
and then for each $k=2,3,\ldots, n-1$,
perform $U_k\otimes U_k$ with probability $1/2$,
while applying the identity operation
with probability $1/2$, respectively.
Here, we remark that $U_k\otimes U_k=I \otimes U_k$.
We now perform $U_{\pi/2}\otimes U_{\pi/2}$ with probability $1/2$ as in \Eref{eq:U_piU_pi},
and then perform the swap operator $U_{01}:\ket{0}\leftrightarrow\ket{1}$ ($\ket{j}\mapsto\ket{j}$ for $2\le j\le n-1$)
with probability $1/2$.
Then we obtain a state of the following form:
\begin{eqnarray}
&\sum_{j=2}^{n-1}& a_j(\ket{0j}\bra{0j}+\ket{1j}\bra{1j}) \nonumber \\
&&+b\left(\ket{\phi^+}\bra{\phi^+}+\ket{\phi^-}\bra{\phi^-}\right)\nonumber \\
&&+c_+\ket{\psi^+}\bra{\psi^+} + c_- \ket{\psi^-}\bra{\psi^-}.\label{eq:diagonalizing}
\end{eqnarray}
Let $T$ be the unitary operator defined by
$\ket{0}\mapsto\ket{0}$, $\ket{1}\mapsto\ket{1}$, $\ket{2}\mapsto\ket{3}$, $\ket{3}\mapsto\ket{4}$,
$\ldots$, $\ket{n-2}\mapsto\ket{n-1}$, and $\ket{n-1}\mapsto\ket{2}$.
Now we perform the following operation:
\[
\rho\mapsto \frac{1}{n-2}\sum_{j=0}^{n-3}(T^j\otimes T^j) \rho (T^j\otimes T^j)^{\dagger}.
\]
Here, we also remark that $T^j\otimes T^j=I \otimes T^j$ for any $j=0,1,\ldots, n-3$.
Then a state in \Eref{eq:diagonalizing} now has the form
\begin{eqnarray}
a& \sum_{i=0}^1&\sum_{j=2}^{n-1}\ket{ij}\bra{ij}
+b\left(\ket{\phi^+}\bra{\phi^+}+\ket{\phi^-}\bra{\phi^-}\right)\nonumber\\
&&+c_+\ket{\psi^+}\bra{\psi^+} + c_- \ket{\psi^-}\bra{\psi^-}.\nonumber
\end{eqnarray}
Let $H$ be the unitary operator defined as
$\ket{0}\mapsto (\ket{0}+\ket{1})/\sqrt{2}$, $\ket{1}\mapsto (\ket{0}-\ket{1})/\sqrt{2}$,
and $\ket{j}\mapsto \ket{j}$ for $2\le j \le n-1$.
After performing as follows:
\[
\rho\mapsto \frac{2}{3}(H\otimes H) \rho (H\otimes H)+ \frac{1}{3}\rho,
\]
let us perform the sequence of the previous operations again.
Then one can easily check that one of states with two parameters in \Eref{eq:2parameter} is obtained,
that is,
there exist unitary operators $U_k$ and probabilities $p_k$
satisfying \Eref{eq:LOCC},
and that furthermore if a state $\rho$ is given by
\begin{eqnarray}
\rho&=&\sum_{i=0}^1\sum_{j=2}^{n-1} a_{ij}\ket{ij}\bra{ij}
+b_+\ket{\phi^+}\bra{\phi^+}+b_-\ket{\phi^-}\bra{\phi^-}\nonumber \\
&&+c_+\ket{\psi^+}\bra{\psi^+} + c_- \ket{\psi^-}\bra{\psi^-}+\cdots\nonumber
\end{eqnarray}
then the two-parameter state transformed by the above procedure
becomes $\rho_{(\alpha,\gamma)}$ in \Eref{eq:2parameter}
where $\alpha=\sum_{i,j}a_{ij}/(2n-4)$ and $\gamma=c_-$.

Noting that $(U\otimes I)\ket{\psi^-}=(I\otimes \pm U)\ket{\psi^-}$
for any $U$ in the group $\mathrm{G}(2,n)$ defined in \Eref{eq:G(m,n)},
we can readily show that $\rho_{(\alpha,\gamma)}$ is invariant under all $U\otimes U$,
that is, for any $U\in \mathrm{G}(2,n)$,
\begin{equation}
(U\otimes U)\rho_{(\alpha,\gamma)}(U^{\dagger}\otimes U^{\dagger})
=\rho_{(\alpha,\gamma)}.\label{eq:UU_invariant}
\end{equation}
We now define the ($U\otimes U$)-{\em twirling} superoperator $\mathcal{T}$ as
\[
\mathcal{T}(\rho)=\int_{\mathrm{G}(2,n)}d\mu_U (U\otimes U)\rho(U^{\dagger}\otimes U^{\dagger}).
\]
and $d\mu_U$ is the normalized Haar measure on $G(2,n)$.
Then it follows from \Eref{eq:LOCC} and \Eref{eq:UU_invariant}
that $\mathcal{T}(\rho)=\rho_{(\alpha,\gamma)}$ for some $\alpha$ and $\gamma$,
and that $\mathcal{T}(\rho_{(\alpha,\gamma)})=\rho_{(\alpha,\gamma)}$, respectively.
We remark that the negativity and the entanglement of formation are entanglement monotones~\cite{VidalW,Vidal}.
Thus, it follows that the negativity and the entanglement of formation for a given $\rho$
are not less than those for $\mathcal{T}(\rho)$, respectively.

\section{Entanglement for the states with two parameters}\label{Sec:Entanglement}
In this section, we consider two measures of entanglement for $\rho_{(\alpha,\gamma)}$,
the negativity and the entanglement of formation,
and explicitly calculate the value of its negativity and bounds on its entanglement of formation.

\begin{figure}
\begin{center}
\epsfig{file=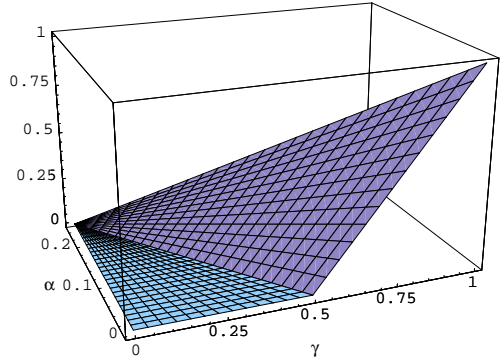, width=.5\linewidth,clip=0.1cm}
\caption{\label{Fig:negativity}The negativity of $\rho_{(\alpha,\gamma)}$ in $2\otimes 4$ quantum system.}
\end{center}
\end{figure}

\begin{figure}
\begin{center}
\epsfig{file=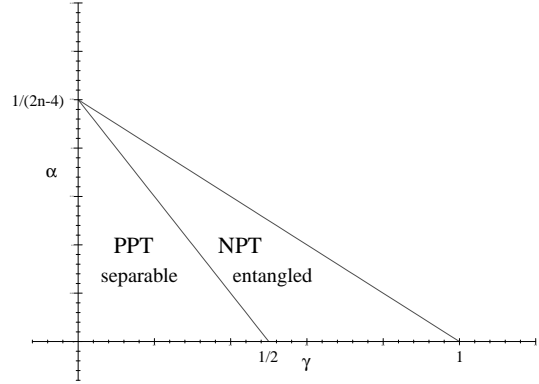, width=.4\linewidth,angle=270,bb=80 84 530 690, clip=1cm}
\caption{\label{Fig:domain}The domain of the parameters $\alpha$ and
$\gamma$ for the states $\rho_{(\alpha,\gamma)}$: All states in
the PPT region are separable and undistillable, and all states in
the NPT region are entangled and distillable.}
\end{center}
\end{figure}

In order to calculate the negativity of $\rho_{(\alpha,\gamma)}$,
we have to compute its partial transpose:
\begin{eqnarray}
\rho_{(\alpha,\gamma)}^{T_B}&=
&\alpha \sum_{i=0}^1\sum_{j=2}^{n-1}\ket{ij}\bra{ij}\nonumber \\
&&+\frac{\beta+\gamma}{2}\left(\ket{01}\bra{01}+\ket{10}\bra{10}+\ket{\phi^-}\bra{\phi^-}\right) \nonumber \\
&&+\frac{3\beta-\gamma}{2}\ket{\phi^+}\bra{\phi^+}.\nonumber
\end{eqnarray}
Since $(3\beta-\gamma)/2=1/2-(n-2)\alpha-\gamma\ge 0$ if and only if
$\rho_{(\alpha,\gamma)}^{T_B}$ is positive,
the negativity of $\rho_{(\alpha,\gamma)}$
 is $\max\{(2n-4)\alpha+2\gamma-1,0\}$,
 whose graph is shown in \Fref{Fig:negativity}.
Then the domain of the parameters $\alpha$ and $\gamma$ for $\rho_{(\alpha,\gamma)}$
consists of two triangular regions, the PPT region satisfying $0\le(n-2)\alpha+\gamma\le 1/2$
and the nonpositive partial transposition (NPT) region satisfying $1/2<(n-2)\alpha+\gamma\le 1$,
as shown in \Fref{Fig:domain}.
We note that all states in the PPT region are separable
since three states, $\rho_{(0,0)}$, $\rho_{(0,1/2)}$, and $\rho_{(1/(2n-4),0)}$,
corresponding to the vertices of the PPT region are separable
and all states in the region are convex combinations of the three states.

\begin{figure}
\begin{center}
\epsfig{file=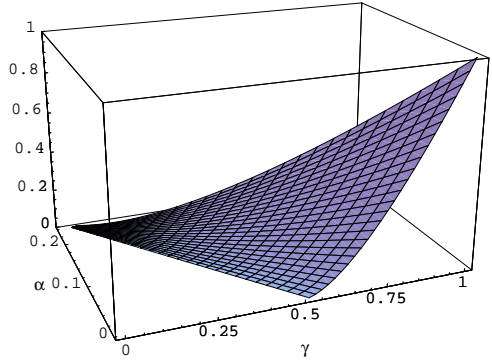, width=.5\linewidth,clip=0.1cm}
\caption{\label{Fig:lowerbound}The lower bound (\ref{Eq:lowerbound2})
on the entanglement of formation of $\rho_{(\alpha,\gamma)}$
for $1/2\le(n-2)\alpha+\gamma\le 1$ in $2\otimes 4$ quantum system.}
\end{center}
\end{figure}

We now consider the entanglement of formation for $\rho_{(\alpha,\gamma)}$.
Even though
it has not been known whether the explicit formula of the entanglement of formation for states
in $2\otimes n$ quantum system can be computed or not,
one can readily
compute one of its lower bounds~\cite{CLLH,Gerjuoy},
\begin{equation}
\mathcal{E}\left(\sqrt{\sum_{i<j}C_{ij}^2} \right),
\label{Eq:lowerbound}
\end{equation}
where
\[
\mathcal{E}(c)=h\left(\frac{1+\sqrt{1-c^2}}{2}\right),
\]
$C_{ij}$ is the Wootters' cuncurrence~\cite{Wootters}
in $2\otimes 2$ dimensional subsystem which is supported by the bases
$\ket{0i}$, $\ket{1i}$, $\ket{0j}$, and $\ket{1j}$.
It is straightforward to check that the lower bound in \Eref{Eq:lowerbound} is
\begin{eqnarray}
\mathcal{E}\left(\mathcal{N}(\rho_{(\alpha,\gamma)})\right)
&=&h\left(\frac{1+\sqrt{1-\mathcal{N}(\rho_{(\alpha,\gamma)})^2}}{2}\right)\nonumber\\
&=&h\left(\frac{1}{2}+\sqrt{((n-2)\alpha+\gamma)(1-(n-2)\alpha-\gamma)}\right)
\label{Eq:lowerbound2}
\end{eqnarray}
since $C_{01}=\mathcal{N}(\rho_{(\alpha,\gamma)})$
and $C_{ij}=0$ unless $i=0$ and $j=1$.
For $2\otimes 4$ quantum system,
the lower bound on 
$E_f\left(\rho_{(\alpha,\gamma)}\right)$
is plotted in \Fref{Fig:lowerbound}.

In order to obtain a tight upper bound of $E_f\left(\rho_{(\alpha,\gamma)}\right)$,
we consider two-parameter states satisfying $2(n-2)\alpha +\gamma=1$, that is,
the states corresponding to the line through two points $(1,0)$ and $(0,1/(2n-4))$ in \Fref{Fig:domain}.
Then the states are of the following form:
\begin{eqnarray}
\rho_{(\alpha,\gamma)}
&=& \alpha\sum_{i=0}^1\sum_{j=2}^{n-1}\ket{ij}\bra{ij}+\gamma\ket{\psi^-}\bra{\psi^-}\nonumber \\
&=& \frac{1-\gamma}{2(n-2)}\sum_{i=0}^1\sum_{j=2}^{n-1}\ket{ij}\bra{ij}+\gamma\ket{\psi^-}\bra{\psi^-}\nonumber\\
&\equiv & \varrho_\gamma.
\label{eq:beta0}
\end{eqnarray}
We note that \Eref{eq:beta0} is an eigenvalue decomposition of a state $\varrho_{\gamma}$,
and that
\begin{equation}
E_f\left(\varrho_{\gamma}\right)\le \gamma
\label{eq:less_than_gamma}
\end{equation}
in view of the convexity of $E_f$.
Let $\varrho_{\gamma}=\sum_{k}p_k\ket{\phi_k}\bra{\phi_k}$ be an optimal decomposition for its entanglement of formation
when $0<\gamma<1$.
By Hughston, Jozsa and Wootters' theorem~\cite{HJW},
there exists a $K\times K$ unitary matrix $U$,
$K$ being greater than or equal to $2n-3$, the rank of $\varrho_\gamma$,
such that for each $k=0,1,\ldots,K-1$
\begin{eqnarray}
\sqrt{p_k}\ket{\phi_k}
=\sum_{i=0}^{1}\sum_{j=2}^{n-1}U_{k,(ij)}^*\sqrt{\frac{1-\gamma}{2(n-2)}}\ket{ij}
+U_{k,2n-4}^*\sqrt{\gamma}\ket{\psi^-},
\label{eq:HJW}
\end{eqnarray}
where $(ij)=i(n-2)+(j-2)$.
Thus it can be obtained from \Eref{eq:HJW} that
\begin{eqnarray}
p_k \tr_B\left(\ket{\phi_k}\bra{\phi_k}\right)
&=&
\sum_{i,i'=0}^{1}\sum_{j=2}^{n-1}U_{k,(ij)}^*U_{k,(i'j)}\frac{1-\gamma}{2(n-2)}
\ket{i}\bra{i'}\nonumber \\
&&+\frac{1}{2}\left|U_{k,2n-4}\right|^2\gamma\left(\ket{0}\bra{0}+\ket{1}\bra{1}\right)
\nonumber \\
&=&
\frac{1-\gamma}{2(n-2)}
\sum_{j=2}^{n-1}\ket{\Psi_{kj}}\bra{\Psi_{kj}}
\nonumber \\
&&+\frac{1}{2}\left|U_{k,2n-4}\right|^2\gamma\left(\ket{0}\bra{0}+\ket{1}\bra{1}\right),
\label{eq:optimal_decomp1}
\end{eqnarray}
where
\[
\ket{\Psi_{kj}}=\sum_{i=0}^{1}U_{k,(ij)}^*\ket{i}.
\]
Then it follows from \Eref{eq:optimal_decomp1} and the concavity of $\mathcal{S}$
that
\begin{eqnarray}
p_k E\left(\ket{\phi_k}\right)
&=&p_k\mathcal{S}\left(\tr_B\ket{\phi_k}\bra{\phi_k}\right)\nonumber\\
&\ge &
\left|U_{k,2n-4}\right|^2\gamma~ \mathcal{S}\left(\frac{1}{2}\left(\ket{0}\bra{0}+\ket{1}\bra{1}\right)\right)
\nonumber \\
&=&\left|U_{k,2n-4}\right|^2\gamma,\label{eq:concave_S}
\end{eqnarray}
since the von Neumann entropy of a pure state is vanished.
Hence, from \Eref{eq:concave_S} and the unitarity of $U$,
we obtain the following inequality:
\begin{eqnarray}
E_f(\varrho_\gamma)
&=&\sum_{k=0}^{K-1}p_k E\left(\ket{\phi_k}\right) \nonumber\\
&\ge &
\sum_{k=0}^{K-1}\left|U_{k,2n-4}\right|^2\gamma \nonumber\\
&=& \gamma.
\label{eq:Eof_S}
\end{eqnarray}
By two inequalities (\ref{eq:less_than_gamma}) and (\ref{eq:Eof_S})
we conclude that $E_f(\varrho_\gamma)=\gamma$.

We now consider the states corresponding to $(\alpha,\gamma)$ in the interior of the NPT region,
that is, $(\alpha,\gamma)$ satisfying $1/2<(n-2)\alpha+\gamma<1$.
By virtue of the convexity of the entanglement of formation,
for a given $(\alpha,\gamma)$ the following inequality can 
be obtained:
\begin{eqnarray}
E_f\left(\rho_{(\alpha,\gamma)}\right)
&=& E_f\left(\rho_{\left(t[1-N_{(\alpha,\gamma)}]/(n-2),t[2N_{(\alpha,\gamma)}-1]+(1-t)N_{(\alpha,\gamma)}\right)}\right)
\nonumber\\
&\le &
t\cdot E_f\left(\rho_{\left([1-N_{(\alpha,\gamma)}]/(n-2),2N_{(\alpha,\gamma)}-1\right)}\right)\nonumber\\
&&+(1-t)\cdot E_f\left(\rho_{\left(0,N_{(\alpha,\gamma)}\right)}\right)\nonumber\\
&=& t\cdot [2N_{(\alpha,\gamma)}-1] + (1-t)\cdot \mathcal{E}(2N_{(\alpha,\gamma)})
\label{eq:upperbound1}
\end{eqnarray}
where $N_{(\alpha,\gamma)}=(n-2)\alpha+\gamma$, and
$t$ is chosen by
\begin{eqnarray}
\alpha
&=&t\cdot[1-N_{(\alpha,\gamma)}]/(n-2),\nonumber\\
\gamma
&=&t\cdot[2N_{(\alpha,\gamma)}-1]+(1-t)\cdot N_{(\alpha,\gamma)}.
\label{eq:t}
\end{eqnarray}

\begin{figure}
\begin{center}
\epsfig{file=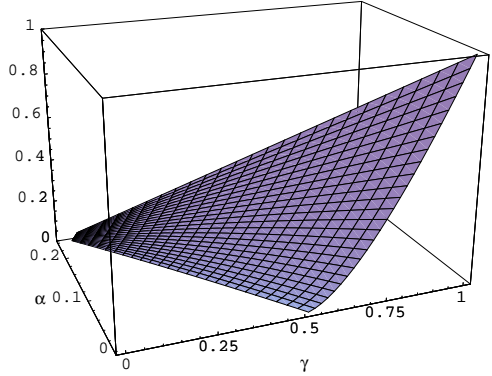, width=.5\linewidth,clip=0.1cm}
\caption{\label{Fig:upperbound}The upper bound (\ref{eq:upperbound2})
on the entanglement of formation of $\rho_{(\alpha,\gamma)}$
for $1/2\le(n-2)\alpha+\gamma\le 1$ in $2\otimes 4$ quantum system.}
\end{center}
\end{figure}

Choosing the appropriate $t$ satisfying \Eref{eq:t},
we are straightforwardly able to calculate
the following upper bound on $E_f\left(\rho_{(\alpha,\gamma)}\right)$:
\begin{eqnarray}
\mathcal{E}\left(\mathcal{N}(\rho_{(\alpha,\gamma)})\right)
+
(n-2)\alpha\frac{2(n-2)\alpha+2\gamma-1-\mathcal{E}
\left(\mathcal{N}(\rho_{(\alpha,\gamma)})\right)}
{1-(n-2)\alpha-\gamma}.\label{eq:upperbound2}
\end{eqnarray}
For $2\otimes 4$ quantum system
the upper bound (\ref{eq:upperbound2}) on $E_f\left(\rho_{(\alpha,\gamma)}\right)$
is shown in \Fref{Fig:upperbound}.
We remark that
the entanglement of formation and its upper bound (\ref{eq:upperbound2})
have the same values
for the states corresponding to three edges of the NPT region in \Fref{Fig:domain}.
Therefore,
we cannot derive any upper bound tighter than the upper bound (\ref{eq:upperbound2})
from the method using the convexity of the entanglement of formation as in \Eref{eq:upperbound1}.
From this viewpoint,
we can say that
the upper bound (\ref{eq:upperbound2})
is a tight upper bound on $E_f(\rho_{(\alpha,\gamma)})$.

We note that $E_f(\varrho_\gamma)=\gamma=\mathcal{N}(\varrho_\gamma)$ and
that $E_f\left(\rho_{(0,\gamma)}\right)\le \mathcal{N}\left(\rho_{(0,\gamma)}\right)$ for all $0\le \gamma\le 1$.
Thus, it follows that
the entanglement of formation of any $\rho_{(\alpha,\gamma)}$ cannot exceed its negativity,
that is,
\[
E_f\left(\rho_{(\alpha,\gamma)}\right)\le \mathcal{N}\left(\rho_{(\alpha,\gamma)}\right),
\]
for any $\rho_{(\alpha,\gamma)}$.

\section{Conclusions}\label{Sec:Conclusion}
In this paper we exhibited a two-parameter class of states $\rho_{(\alpha,\gamma)}$
in $2\otimes n$ quantum system, found that an arbitrary state can be transformed into
$\rho_{(\alpha,\gamma)}$ by means of LOCC,
and showed that $\rho_{(\alpha,\gamma)}$ is invariant under unitary operations of the form $U\otimes U$
on $2\otimes n$ quantum system.
We finally investigated the entanglement for $\rho_{(\alpha,\gamma)}$,
by computing two measures of entanglement, the negativity and the entanglement of formation.

For a higher dimensional quantum system, that is, an $m\otimes n$ quantum system ($m<n$),
we can exhibit a two-parameter class given by
\begin{eqnarray}
\alpha &&\sum_{i=0}^{m-1}\sum_{j=m}^{n-1}\ket{ij}\bra{ij} \nonumber \\
&&+\beta \left(\sum_{i,j=0(i<j)}^{m-1}\ket{\varphi_{ij}^+}\bra{\varphi_{ij}^+}
+\sum_{k=0}^{m-1}\ket{kk}\bra{kk}\right) \nonumber \\
&&+\gamma \sum_{i,j=0(i<j)}^{m-1}\ket{\varphi_{ij}^-}\bra{\varphi_{ij}^-}
\label{eq:higher_2parameter}
\end{eqnarray}
where
\[
\ket{\varphi_{ij}^\pm}=\frac{1}{\sqrt{2}}\left(\ket{ij}\pm\ket{ji}\right),
\]
and
\[
m(n-m)\alpha+\frac{m(m+1)}{2}\beta+\frac{m(m-1)}{2}\gamma=1.
\]
Furthermore, we can show that
any state in $m\otimes n$ quantum system can be transformed
into a state of the form of \Eref{eq:higher_2parameter}
by LOCC,
and that any state in this class is ($U\otimes U$)-invariant for all unitary $U$ in
the group $\mathrm{G}(m,n)$ defined in \Eref{eq:G(m,n)}.
Since every state in this class has properties analogous to those of $\rho_{(\alpha,\gamma)}$,
one could investigate its entanglement.

\ack{
This work was supported by
a Korea Research Foundation Grant~(KRF-2000-015-DP0031)
and
a KIAS Research Fund (No. 02-0140-001).
SL would like to thank Prof. Sungdam Oh in Sookmyung Women's University
and Prof. Jaewan Kim in KIAS
for very useful discussions.}

\section*{References}


\begin{thebibliography}{100}

\bibitem{BDSW} Bennett~C~H, DiVincenzo~D~P, Smolin~J~A and Wootters~W~K 1996
{\it Phys. Rev. A} {\bf 54} 3824

\bibitem{Wootters} Wootters~W~K 1998
{\it Phys. Rev. Lett.} {\bf 80} 2245

\bibitem{Horodecki} Horodecki~M 2001
{\it Quantum Inf. Comput.} {\bf 1} 3

\bibitem{Wootters2} Wootters~W~K 2001
{\it Quantum Inf. Comput.} {\bf 1} 27

\bibitem{VDC} Vidal~G, D\"{u}r~W and Cirac~J~I 2002
{\it Phys. Rev. Lett.} {\bf 89} 027901

\bibitem{Peres} Peres~A 1996
{\it Phys. Rev. Lett.} {\bf 77} 1413.

\bibitem{Horodeckis1} Horodecki~M, Horodecki~P and Horodecki~R 1996
{\it Phys. Lett. A} {\bf 223} 1

\bibitem{VidalW} Vidal~G and Werner~R~F 2002
{\it Phys. Rev. A} {\bf 65} 032314

\bibitem{Vidal} Vidal~G 2000
{\it J. Mod. Opt.} {\bf 47} 355

\bibitem{Horodecki1} Horodecki P 1997
{\it Phys. Lett. A} {\bf 232} 333

\bibitem{DCLB} D\"{u}r~W, Cirac~J~I, Lewenstein~M and Bru\ss~D 2000
{\it Phys. Rev. A} {\bf 61} 062313

\bibitem{Horodeckis2} Horodecki~M, Horodecki~P and Horodecki~R 1998
{\it Phys. Rev. Lett.} {\bf 80} 5239

\bibitem{SZ} Scully~M~O and Zubairy~M~S 1999
{\em Quantum Optics} (Cambridge: Cambridge University Press)

\bibitem{HW}
 Hill~S and Wootters~W~K 1997
{\it Phys. Rev. Lett.} {\bf 78} 5022

\bibitem{TV} Terhal~B~M and Vollbrecht~K~G~H. 2000
{\it Phys. Rev. Lett.} {\bf 85}, 2625

\bibitem{VollbrechtW} Vollbrecht~K~G~H and Werner~R~F 2001
{\it Phys. Rev. A} {\bf 64} 062307

\bibitem{CLLH}
Chen~P, Liang~L, Li~C and Huang~M 2002
{\it Phys. Lett. A} {\bf 295} 175

\bibitem{Gerjuoy}
Gerjuoy~E 2003
{\it Phys. Rev. A} {\bf 67} 052308

\bibitem{Werner} Werner~R~F, 1989
{\it Phys. Rev. A} {\bf 40} 4277

\bibitem{HJW} Hughston~L~P, Jozsa~R and Wootters~W~K 1993
{\it Phys. Lett. A} {\bf 183} 14



\end{thebibliography}
\end{document}